\documentclass{article}
\usepackage{fullpage}
\DeclareMathAlphabet{\altmathcal}{OMS}{cmsy}{m}{n}
\usepackage{mathptmx}
\usepackage{color}
\usepackage{amsmath,amsbsy}
\usepackage{amsfonts,amssymb}
\usepackage[square,numbers,sort&compress,super]{natbib}
\usepackage{url}
\usepackage{bm}
\usepackage{mdwlist}
\usepackage{float}
\usepackage{graphicx}
\usepackage{subcaption}
\usepackage{fancyhdr}
\usepackage{hyperref}
\usepackage{dsfont}

\title{Uncertainty quantification for epidemiological forecasts of COVID-19 through combinations of model predictions}
\author{D.~S.~Silk$^\dagger$, V.~E.~Bowman$^\dagger$, D. Semochkina$^{\ddagger}$, U. Dalrymple$^\dagger$ and D.~C.~Woods$^{\ddagger}$  \\
$^\dagger$Defence Science and Technology Laboratory, Porton Down \\
$^\ddagger$Statistical Sciences Research Institute, University of Southampton}

\begin{document}

\maketitle

\thispagestyle{fancy}

Scientific advice to the UK government throughout the COVID-19 pandemic has been informed by ensembles of epidemiological models provided by members of the Scientific Pandemic Influenza group on Modelling (SPI-M). Among other applications, the model ensembles have been used to forecast daily incidence, deaths and hospitalizations. The models differ in approach (e.g. deterministic or agent-based) and in assumptions made about the disease and population. These differences capture genuine uncertainty in the understanding of disease dynamics and in the choice of simplifying assumptions underpinning the model. Although analyses of multi-model ensembles can be logistically challenging when time-frames are short, accounting for structural uncertainty can improve accuracy and reduce the risk of over-confidence in predictions. In this study, we compare the performance of various ensemble methods to combine short-term (14 day) COVID-19 forecasts within the context of the pandemic response. We address practical issues around the availability of model predictions and make some initial proposals to address the short-comings of standard methods in this challenging situation.

%
%
%
%
%
%
%

\section{Introduction}

Comprehensive uncertainty quantification in epidemiological modelling is a timely and challenging problem. During the COVID-19 pandemic, a common requirement has been statistical forecasting in the presence of an ensemble of multiple candidate models. For example, multiple candidate models may be available to predict disease case numbers, resulting from different modelling approaches (e.g. mechanistic or empirical) or differing assumptions about spatial or age mixing. Alternative models capture genuine uncertainty in scientific understanding of disease dynamics, different simplifying assumptions underpinning each model derivation, and/or different approaches to the estimation of model parameters. While the analysis of multi-model ensembles can be computationally challenging, accounting for this `structural uncertainty' can improve forecast accuracy and reduce the risk of over-confident prediction \cite{semenov2010use,raftery1997bayesian}. 
 
A common ensemble approach is model averaging, which tries to find an optimal combination of models in the space spanned by all individual models \cite{madigan1996bayesian,hoeting1999bayesian}.  However, in many settings this approach can fail dramatically: (i) the required marginal likelihoods (or equivalently Bayes factors) can depend on arbitrary specifications for non-informative prior distributions of model-specific parameters; and (ii) asymptotically the posterior model probabilities, used to weight predictions from different models, converge to unity on the model closest to the `truth'. While this second feature may be desirable when the set of models under consideration contains the true model (the $\altmathcal{M}$-closed setting), it is less desirable in more realistic cases when the model set does not contain the true data generator ($\altmathcal{M}$-complete and $\altmathcal{M}$-open). Here, this property of Bayesian model averaging asymptotically choosing a single model can be thought of as a form of overfitting. For these latter settings, alternative methods of combining predictions from model ensembles may be preferred, for example, via combinations of individual predictive densities \cite{wallis}. Combination weights can be chosen via application of predictive scoring, as commonly applied in meteorological and economic forecasting \cite{gneiting2005,mcdonald2011}.  

If access to full posterior information is available, other approaches are also possible. Model stacking methods \cite{yao2018using}, see Section~\ref{sec:stack}, can be applied directly, ideally using leave-one-out predictions or sequential predictions of future data to mitigate over-fitting. Alternatively, the perhaps confusingly named Bayesian model combination method \cite{minka2000bayesian,monteith2011turning} could be employed, where the ensemble is expanded to include linear combinations of the available models. For computationally expensive models, where straightforward generation of predictive densities is prohibitive, a statistical emulator built from a Gaussian process priors \cite{kennedy2001bayesian} could be assumed for each expensive model. Stacking or model combination can then be applied to the resulting posterior predictive distributions, conditioning on model runs and data. 

In this paper we explore the process of combining short-term epidemiological forecasts for COVID-19 daily deaths, and hospital and intensive care unit (ICU) occupancy, within the context of supporting UK decision makers during the pandemic response. In practice, this context placed constraints on the information available to the combination algorithms. In particular, the individual model posterior distributions were unavailable, which prohibited the use of the preferred approach outlined above, and so alternative methods had to be utilized. The construction of consensus forecasts in the UK has been undertaken through a mixture of algorithmic model combination and expert curation by the Scientific Pandemic Influenza group on Modelling (SPI-M). For time-series forecasting, equally-weighted mixture models have been employed \cite{sage2020b,sage2020c,sage2020}. We compare this approach to more sophisticated ensemble methods. Important related work is the nowcasting of the current state of the disease within the UK population through metrics such as the effective reproduction number, growth rate and doubling time \cite{maishman2021}. 

The rest of the paper is organized as follows. Section~\ref{Combinations} describes in more detail methods of combining individual model predictions. Limitations of the available forecast data for the COVID-19 response is then described in Section~\ref{COVID-19}, and Section~\ref{results} compares the performance of ensemble algorithms and individual model predictions. Section~\ref{discussion} provides some discussion and areas for future work. This paper complements work undertaken in a wider effort to improve the policy response to COVID-19, in particular a parallel effort to assess forecast performance \cite{seb2020}. 

\section{Combinations of predictive distributions}\label{Combinations}

Let $\bm{y}=(y_1,\ldots, y_n)^\intercal$ represent the observed data with $\altmathcal{M} = \left(M_1,\ldots, M_K\right)$ an ensemble of models, with the $k$th model having posterior predictive density $p_k(\tilde{y} \mid \bm{y})$, where $\tilde{y}$ is the future data. We consider two categories of ensemble methods: (i) those that stack the predictive densities as weighted mixture distributions (also referred to as pooling in decision theory); and (ii) those that use regression models to combine point predictions obtained from the individual posterior predictive densities. In both cases, stacking and regression weights can be chosen using scoring rules.

\subsection{Scoring rules}\label{sec:score}

Probabilistic forecast quality is often assessed via a scoring rule $S(p,y)\in {\rm I\!R}$ \cite{gneiting2007,gneiting2011} with arguments $p$, a predictive density, and $y$, a realisation of future outcome $Y$. Throughout, we apply negatively-orientated scoring rules, such that a lower score denotes a better forecast. A proper scoring rule ensures that the minimum expected score is obtained by choosing the data generating process as the predictive density. That is, if $d$ is the density function from the true data generating process, then
$$\mathbb{E}_d(S(d,Y))=\int d(y)S(d,y)\;\mathrm{d} y\leq \int d(y)S(p,y)\;\mathrm{d}y  =\mathbb{E}_d(S(p,Y))\,,$$
for any predictive density $p$. Common scoring rules include: 
\begin{enumerate}
\item log score: $S_l(p, y) = -\log p(y)$;
\item continuous ranked probability score (CRPS): $S_c(p, y) = \mathbb{E}_p(Y - y)-\frac{1}{2}\mathbb{E}_p(Y - Y^\prime) $, with $Y, Y^\prime\sim p$ and having finite first moment. 
\suspend{enumerate}
For deterministic predictions,  i.e. $p$ being a point mass density with support on $x$, CRPS reduces to the mean absolute error $S(x, y) = |x - y|$, and hence this score can be used to compare probabilistic and deterministic predictions. If only quantiles from $p$ are available, alternative scoring rules include
\resume{enumerate}
\item quantile score: $S_{q,\alpha}(p, y) = (\mathds{1}\{y<q\} - \alpha)(q - y)$ for quantile forecast $q$ from density $p$ at level $\alpha \in (0, 1)$; 
\item interval score: $S_{I, \alpha}(p, y) =  (u-l) + \frac{2}{\alpha}(l - y)\mathds{1}\{y<l\} + \frac{2}{\alpha}(y - u)\mathds{1}\{y>u\}$ for$(l,u)$ being a central $(1-\alpha)\times 100$\% prediction interval from $p$.    
\end{enumerate}
Quantile and interval scores can be averaged across available quantiles/intervals to provide a score for the predictive density. The CRPS is the integral of the quantile score with respect to $\alpha$. Scoring rules can be used to rank predictions from individual models or to combine models from an ensemble, for example using stacking or regression methods.

\subsection{Stacking methods}\label{sec:stack}

Given an ensemble of models $\altmathcal{M}$, stacking methods result in a posterior predictive density of the form
\begin{equation*}\label{mix}
p(\tilde{y} \mid \bm{y}) = \sum_{k=1}^K w_kp_k(\tilde{y} \mid \bm{y})\,,
\end{equation*}
where $p_k$ is the (posterior) predictive density from model $M_k$ and $w_k\ge 0$ weights the contribution of the $k$th model to the overall prediction, with $\sum_k w_k = 1$. Giving equal weighting to each model in the stack has proved surprisingly effective in economic forecasting \cite{mcdonald2011}. Alternatively, given a score function $S$ and out-of-sample ensemble training data $\tilde{y}_1,\ldots,\tilde{y}_m$, weights $\bm{w} = (w_1, \ldots, w_K)^\intercal$ can be chosen via 
\begin{equation}\label{logscore}
\min_{\bm{w}\in\altmathcal{S}_K} \frac{1}{m}\sum_{i=1}^m S(p(\tilde{y}_i \mid \bm{y}), \tilde{y}_i)\,,
\end{equation}
where $\altmathcal{S}_K = \{\bm{w} \in [0, 1]^K\,:\, \sum_k w_k = 1\}$ is the $K$-dimensional simplex. This approach is the essence of Bayesian model stacking as described by \citet{yao2018using}. 

Alternatively, scoring functions can be used to construct normalized weights 
\begin{equation}\label{aic}
w_k = \frac{f\left(S_k \right)}{\sum_k f\left(S_k\right)}\,,
\end{equation}
\noindent with $S_k =  \sum_i S(p_k(\tilde{y}_i \mid \bm{y}) / m$ being the average score for the $k$th model, and $f$ being an inversely monotonic function; with the log-score~\eqref{logscore} and $f(S) = \exp(-S)$, Akaike Information Criterion (AIC) style weights are obtained\cite{burnham2002}.

%

\subsection{Regression-based methods} \label{section:regression}
Ensemble model predictions can also be formed using regression methods with covariates that correspond to point forecasts from the model ensemble. These methods are particularly suited  to ``low-resolution'' posterior predictive information where only posterior summaries are available and the covariates can be defined directly from, e.g., reported quantiles. We consider two such regression-based methods: Ensemble Model Output Statistics \cite{gneiting2005calibrated} (EMOS) and Quantile Regression Averaging \cite{nowotarski2015computing} (QRA).

EMOS defines the ensemble prediction in the form of a Gaussian distribution
\begin{equation}\label{EMOS}
\hat{y} \sim N(a + b_1\hat{y}_1 + ... + b_K\hat{y}_K, c+d V^2),
\end{equation} 
where $\hat{y}_1,...,\hat{y}_K$ are point forecasts from the individual models, $c$ and $d$ are non-negative coefficients, $V = \sum_k (\hat{y}_k - \bar{y})^2 / (K-1)$ is the ensemble variance with $\bar{y} = \sum_k \hat{y}_k / K$ the ensemble mean, and $a$ and $b_1,...,b_K$ are regression coefficients. Tuning of the coefficients is achieved by minimizing the average CRPS using out-of-sample data 
$$
\bar{S}_c(\hat{y}) = \frac{1}{m}\sum_{i=1}^m S_c(\hat{y}, \tilde{y}_i)\,.
$$
For $\hat{y}$ following distribution~\eqref{EMOS}, $S_c$ is available in closed form \cite{gneiting2005calibrated}. To aid interpretation, the coefficients can be constrained to be non-negative (this version of the algorithm is known as EMOS$+$).  

QRA defines a predictive model for each quantile level, $\beta$, of the ensemble forecast as
\begin{equation}\label{QRA}
\hat{y}(\beta) = b_1\hat{y}_{1}(\beta) + ... + b_K\hat{y}_{K}(\beta),
\end{equation}
where $\hat{y}_{k}(\beta)$ is the $\beta$-level quantile of the (posterior) predictive distribution for model $k$. We make the parsimonious assumption that the non-negative coefficients, $b_1,...b_K$, are independent of the level $\beta$, and estimate them by minimizing the weighted average interval score across $n_{\alpha}$ central $(1-\alpha)\times 100\%$ predictive intervals defined from the quantiles, and $m$ out-of-sample data points: 
$$
\bar{S}_I = \frac{1}{mn_{\alpha}}\sum_{i = 1}^m \sum_{\alpha} \frac{\alpha}{4} S_{I, \alpha}(\hat{y}(\alpha), \tilde{y}_i)\,,
$$
where the factors $\alpha/4$ weight the interval scores such that at the limit of including all intervals, the score approaches the CRPS.

\section{COVID-19 pandemic forecast combinations}\label{COVID-19}




For the COVID-19 response, probabilistic forecasts from multiple epidemiological models were provided by members of SPI-M at weekly intervals. Especially in the early stages of the pandemic, the model structures and parameters were evolving to reflect increased understanding of COVID-19 epidemiology and changes in social restrictions. The impact of evolving models is discussed in Section \ref{discussion}. Practical constraints on data bandwidth and rapid delivery schedules resulted in individual model forecasts being reported as quantiles of the predictive distributions for the forecast window, and minimal information was available about the individual posterior distributions. Whilst EMOS and QRA can be directly applied using only posterior summaries, to implement stacking we estimated posterior densities as skewed-Normal distributions fitted to each set of quantiles. Stacking weights are obtained using~\eqref{aic}, with $f$ taken to be the reciprocal function \cite{mcdonald2011evaluating}, that is 
\begin{equation}\label{reciprocal_weights}
w_k = \frac{\sum_{i=1}^{m}\lambda^{m-i}\bar{S}_{ik}^{-1}}{S_K}
\end{equation}
with 
$$
\bar{S}_{ik} = \sum_\alpha S_{q, \alpha}(\hat{y}_k(\alpha), \tilde{y}_i)\,,
$$
and
$$
S_K = \sum_{k=1}^K{\sum_{i=1}^{m}\lambda^{m-i}\bar{S}_{ik}^{-1}}\,.
$$
The exponential decay term (with $\lambda=0.9$) controls the relative influence of more recent observations.

In Section~\ref{results}, three choices of stacking weights are compared: (i) Reciprocal weights~\eqref{reciprocal_weights} which are invariant with respect to future observation times $t+1, \ldots, t+m$, (ii) equal weights $w_k = 1/K$, and (iii) time-varying weights constructed via exponential interpolation between (i) and (ii) to reduce the influence on forecasts further in the future of the performance of individual models in the training window.

\section{Comparing the performance of ensemble and individual forecasts}\label{results}

The performance of the different ensemble and individual forecasts from $K=9$ models provided by SPI-M were assessed over a set of $14$ day forecast windows. Each model is the evolving output from teams of independent researchers. Individual model predictions were provided for three different COVID-19 related quantities (see Table~\ref{results:valuetypes}) for different UK nations and regions (see Table~\ref{results:nationsregions}) for the $m=20$ days immediately preceding the forecast window. Corresponding ensemble training data were obtained from government provided data streams. The individual model predictions used were the most recently reported that had not been conditioned on data from the combination training window. The assessment was conducted after a sufficient delay such that effects of under reporting on the observational data was negligible. However, it is unknown whether individual SPI-M models attempted to account for potential under-reporting bias in the data used for parameter fitting.

\begin{table}[H] 

      \centering
        \begin{tabular}{ll}
						\hline
						Value types & Description \\ \hline
						death\_inc\_line & New daily deaths by date of death\\
						hospital\_prev & Hospital bed occupancy\\ 
						icu\_prev & Intensive care unit (ICU) occupancy\\ \hline
        \end{tabular}
				\caption{COVID-19 value types (model outputs of interest) for which forecasts were scored.}
				\label{results:valuetypes}
\end{table}

\begin{table}[H] 		

      \centering
        \begin{tabular}{ll}
						\hline
						Nations & Regions \\ \hline
						England & London\\
						Scotland & East of England\\
						Wales & Midlands\\ 
						Northern Ireland & North East and Yorkshire\\
						& North West \\
						& South East \\
						& South West \\ \hline
        \end{tabular}
				\caption{Nations/regions for which forecasts were scored.}
				\label{results:nationsregions}

\end{table}

We present results for the individual models and stacking, EMOS and QRA ensemble methods, as described in Section~\ref{Combinations}. Data-driven, equal and time-varying weights were applied with model stacking, see Section~\ref{COVID-19}. EMOS coefficients were estimated by minimizing CRPS, with the intercept set to zero to force the combination to use the model predictions. While this disabled some potential for bias correction, it was considered important that the combined forecasts could be interpreted as weighted averages of the individual model predictions. QRA was parameterized via minimization of the average of the weighted interval scores for the $0\%$ (i.e. the quantile score for the median), $50\%$ and $90\%$ prediction intervals, as described in Section~\ref{section:regression}. The same score was used to calculate the stacking weights in ~\eqref{reciprocal_weights}. In each case, $m=20$ ensemble training data points were used, and optimization was performed using a particle swarm algorithm \cite{kennedy1995particle}.

Performance was measured for each model/ensemble method using the weighted average of the interval score over the 14 day forecast window and $0\%$,  $50\%$ and $90\%$ intervals. In addition, three well-established assessment metrics were calculated; {\it sharpness}, {\it bias} and {\it calibration} \cite{gneiting2007probabilistic}. Sharpness ($\sigma$) is a measure of prediction uncertainty, and is defined here as the average width of the $50\%$ central prediction interval over the forecast window. Bias ($\beta$) measures over- or under-prediction of a predictive model as the proportion of predictions for which the reported median is greater than the data value over the forecast window. Calibration ($\gamma$) quantifies the statistical consistency between the predictions and data, via the proportion of predictions for which the data lies inside the $50\%$ central predictive interval. The bias and calibration scores were linearly transformed as $\hat{\beta}=(0.5-\beta)/0.5$ and $\hat{\gamma}=(0.5-\gamma)/0.5$, such that a well-calibrated prediction with no bias or uncertainty corresponds to $(\hat{\beta},\hat{\gamma}, \sigma) = (0,0,0)$.

Table~\ref{results:table_values} summarizes the performance of the ensemble methods and averaged performance of the individual models across the four forecast windows. Averaged across nations/regions, the best performing forecasts for new daily deaths were obtained using stacking with time invariant weights; for hospital bed occupancy QRA performed best; and for ICU occupancy, the best method was EMOS. The lowest interval scores occur when predicting new daily deaths, reflecting the more accurate and precise individual model predictions for this output, relative to the others. The ensemble methods also all perform similarly for this output. Importantly, every ensemble method improves upon the average scores for the individual models. This means that with no prior knowledge about model fidelity, if a single forecast is desired, it is better on average to use an ensemble than to select a single model.

\begin{table}[H] 
    \begin{subtable}{.5\textwidth}
        \caption{death\_inc\_line}
        \begin{tabular}{cccc}
						Model & $\bar{S}_{I}$ & $\hat{\beta}$ & $\hat{\gamma}$ \\ 
						\hline
						Stacked: time-invariant weights & 2.16 & 0.66 & 0.49 \\
						Stacked: equal-weights & 2.25 & 0.72 & 0.57  \\
						QRA & 2.28 & 0.65 & 0.51  \\
						Stacked: time-varying weights & 2.43 & 0.76 & 0.45\\
						EMOS & 2.82 & 0.76 & 0.54 \\
						Models & 3.03 & 0.77 & 0.61 \\
        \end{tabular}
    \end{subtable}%
		\begin{subtable}{.5\textwidth}
        \caption{hospital\_prev}
        \begin{tabular}{cccc}
						Model & $\bar{S}_{I}$ & $\hat{\beta}$ & $\hat{\gamma}$ \\ 
						\hline
						QRA & 18.0 & 0.80 & 0.78  \\
				    Stacked: time-invariant weights & 24.5 & 0.81 & 0.68  \\
						Stacked: time-varying weights & 25.4 & 0.84 & 0.77  \\
						EMOS & 25.4 & 0.82 & 0.76  \\		
						Stacked: equal-weights & 27.6 & 0.82 & 0.79  \\
						Models & 33.2 & 0.88 & 0.77 \\
        \end{tabular}
    \end{subtable}
				\begin{subtable}{.5\textwidth}
        \caption{icu\_prev}
        \begin{tabular}{cccc}
						Model & $\bar{S}_{I}$ & $\hat{\beta}$ & $\hat{\gamma}$ \\ 
						\hline
						EMOS & 2.62 & 0.76 & 0.63 \\
						QRA & 3.68 & 0.84 & 0.75  \\
						Stacked: time invariant weights & 3.8 & 0.78 & 0.75  \\
						Stacked: time varying weights & 3.84 & 0.78 & 0.74  \\
						Stacked: equal weights & 4.07 & 0.79 & 0.77  \\
						Models & 4.28 & 0.85 & 0.74 \\

				\end{tabular}
    \end{subtable}
    
		\caption{Median interval and mean bias and calibration scores for each value type, averaged over regions/nations for each ensemble algorithm. The mean score for the individual models is also shown. Rows are ordered by increasing interval score in each subtable. The median was chosen for the interval score due to the presence of extreme values.} \label{results:table_values}
\end{table}

\begin{figure}
	\centering
		\includegraphics[width=0.8\textwidth]{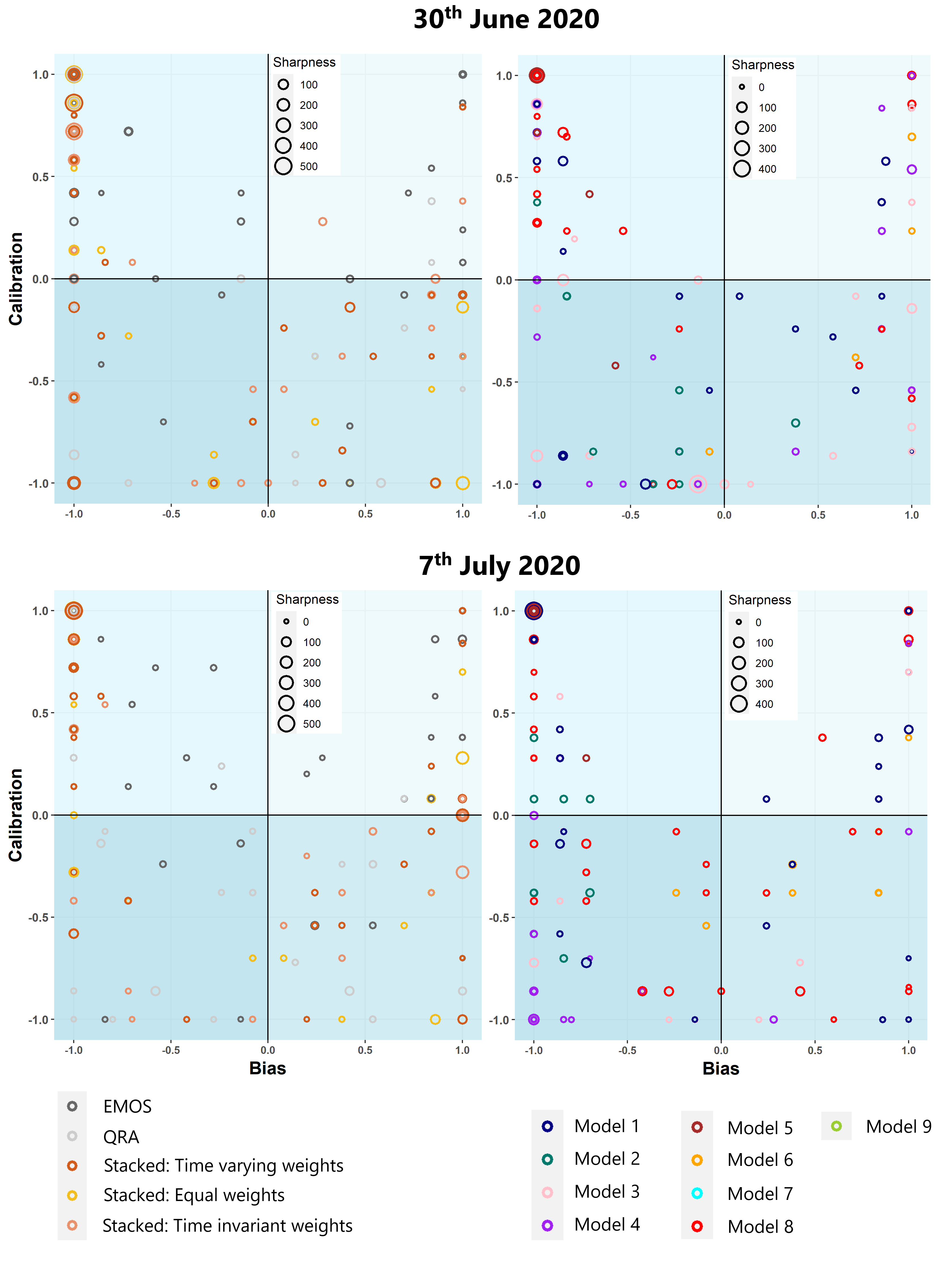}
	\caption{Sharpness, bias and calibration scores for the (left) individual and (right) ensemble forecasts, for all regions and value types delivered on (top) 30th June and (bottom) 7th July, 2020. Note that multiple points are hidden when they coincide. The shading of the quadrants (from darker to lighter) implies a preference for over-prediction rather than under-prediction, and for prediction intervals that contain too many data points, rather than too few.}
	\label{results:allpointscombined20200511}
\end{figure}

\begin{figure}
	\centering
		\includegraphics[width=0.8\textwidth]{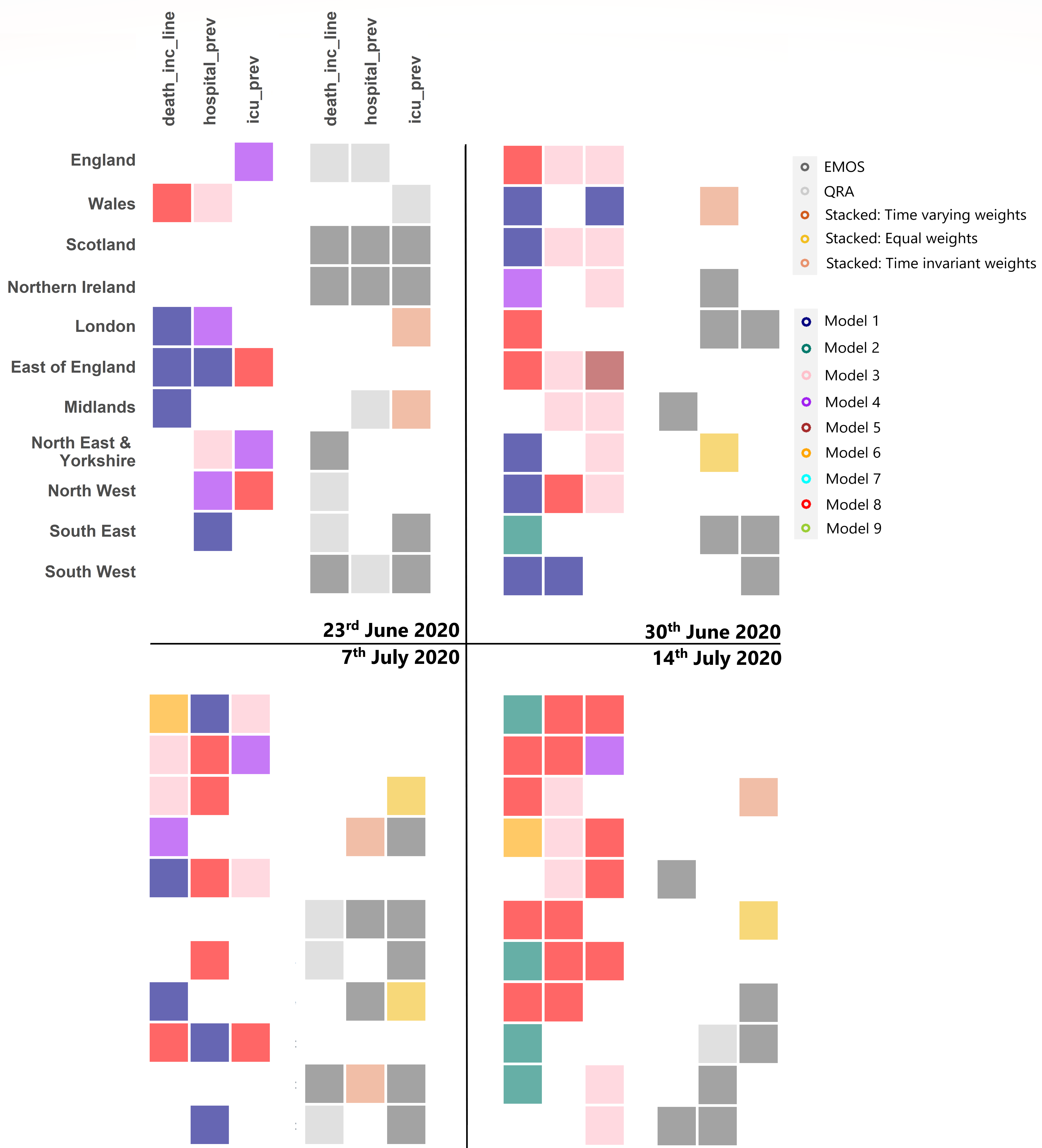}
	\caption{The best performing individual model or ensemble method for each region/nation and value type (for forecasts delivered on the 23rd and 30th June, and 7th and 14th July 2020), evaluated using the absolute distance from the origin on the calibration-bias plots. Ties were broken using the sharpness score. For each date, only the overall best performing model / ensemble is displayed, but for clarity the results are separated into (left) individual models and (right) combinations. Region/nation and value type pairs for which there was less than two individual models with both training and forecast data available were excluded from the analysis.}
	\label{results:individualresults}
\end{figure}

To examine the performance of the ensemble methods, and individual model predictions, in more detail, we plot results for the nations/regions separately for different forecast windows, see Figure~\ref{results:allpointscombined20200511} for two examples. We plot bias ($\hat{\beta}$) against calibration ($\hat{\gamma}$) and divide the plotting region in four quadrants; the top-right quadrant represents perhaps the least worrying errors, as here the methods over-predict the outcomes with prediction intervals that are under-confident. Where both observational data and model predictions were available, the metrics were evaluated for the three value types and eleven regions/nations. Unfortunately, it is difficult to ascertain any patterns for either example. The best forecasting model was also highly variable across nations/regions and value types (as shown for calibration and bias in Figure~\ref{results:individualresults}), and was often an individual model.

\begin{figure}[H]
	\centering
		\includegraphics[width=0.9\textwidth]{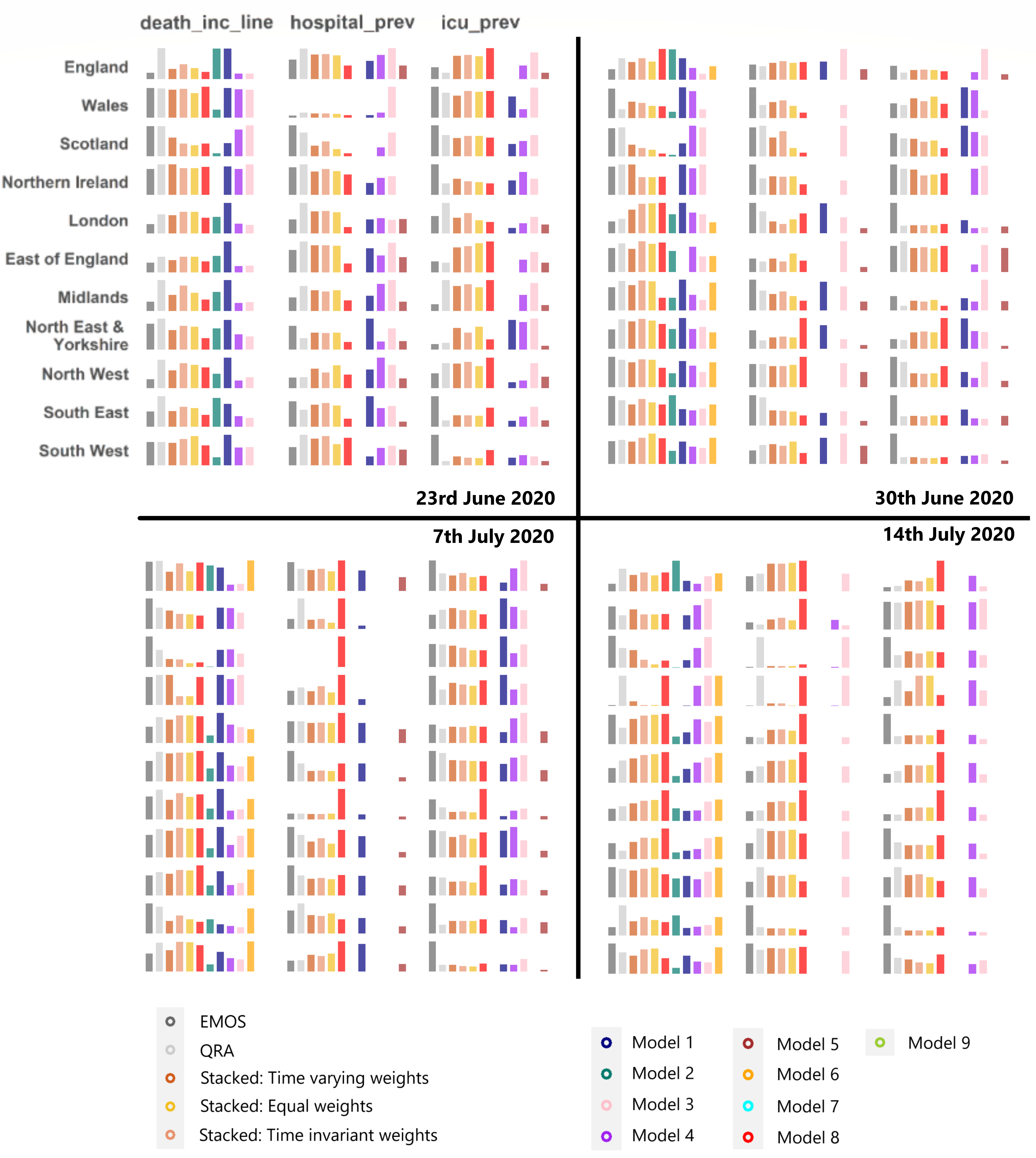}
	\caption{Performance of individual models and ensemble methods for each region/nation and value type (for forecasts delivered on the 23rd and 30th June, and 7th and 14th July 2020). The height of each bar is calculated as the reciprocal of the weighted average interval score, so that higher bars correspond to better performance. Gaps in the results correspond to region/nation and value type pairs for a which a model did not provide forecasts. No combined predictions were produced for Scotland hospital\_prev on the 7th July as forecasts were only provided from a single model.}
	\label{results:individualbars}
\end{figure}

\begin{figure}[H]
	\centering
		\includegraphics[width=0.4\textwidth]{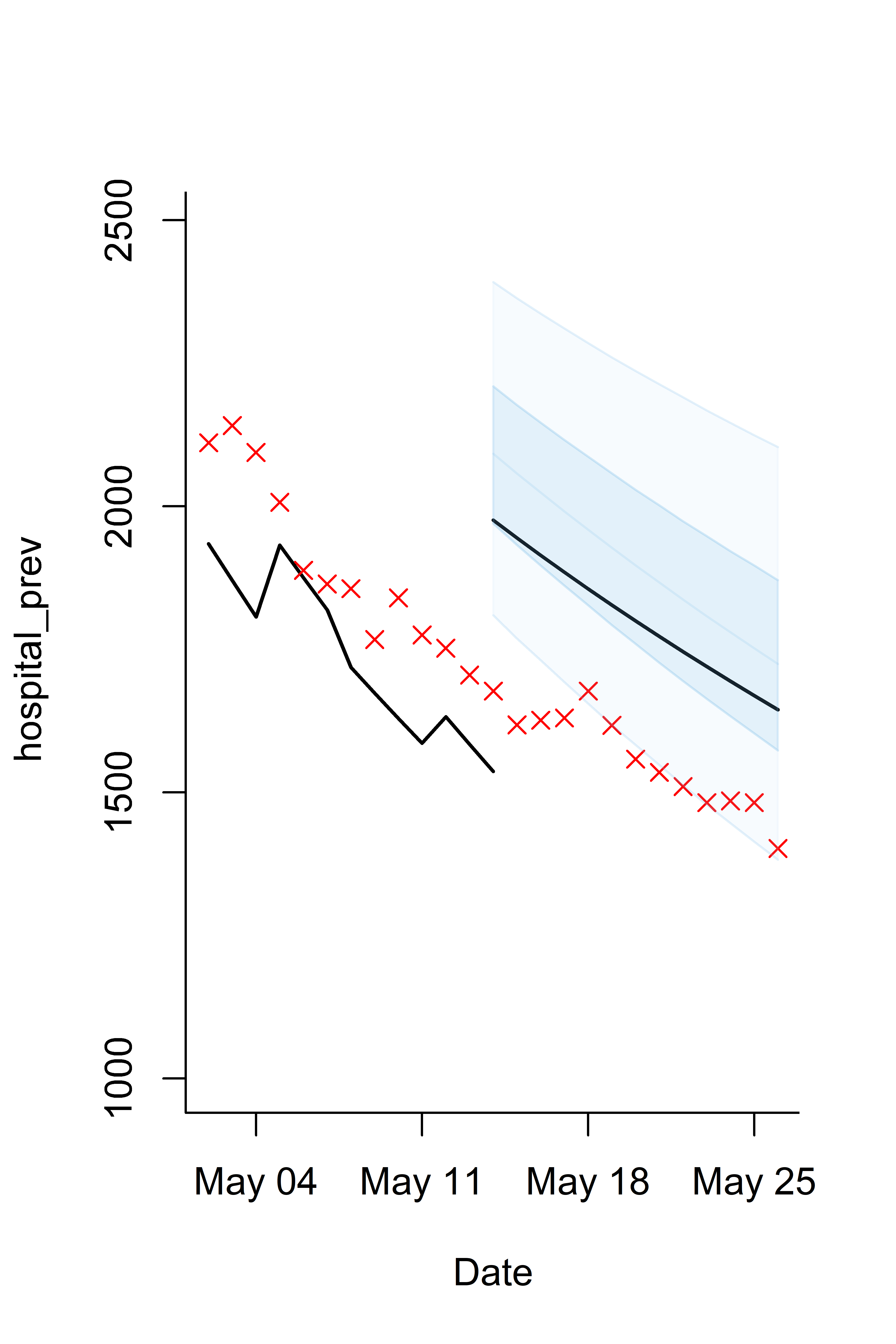}
	\caption{QRA forecast for hospital bed occupancy in the North West region. A large discontinuity between the current and past forecasts (black line) of the individual model corresponding to the covariate with the largest regression coefficient can lead to increased bias for the QRA algorithm. The median, $50\%$ and $90\%$ QRA prediction intervals are shown in blue, while the data is shown in red.}
	\label{results:RegressionProblem}
\end{figure}

The variability in performance was particularly stark for QRA and EMOS, which were found in some cases to vastly outperform the individual and stacked forecasts, and in others to substantially under-perform (as shown in Figure~\ref{results:individualbars}). The former case generally occurred when all the individual model training predictions were highly biased (for example, for occupied ICU beds in the South West for the 23rd June forecast). In these cases, the non-convexity of the QRA and EMOS coefficients led to forecasts that were able to correct for this bias. The problem of bias correction is, of course, the case for which these methods were originally proposed in weather forecasting. Whether this behaviour is desirable depends upon whether the data is believed to be accurate, or itself subject to large systematic biases, such as under-reporting, that is not accounted for within the model predictions. The latter case of under-performance occurred in the presence of large discontinuities between the individual model training predictions and forecasts, corresponding to changes in model structure or parameters. This disruption to the learnt relationships between individual model predictions, and to the data (as captured by the regression models), led to increased forecast bias (an example of this is shown in Figure~\ref{results:RegressionProblem}).

In comparison, the relative performance of the stacking methods was more stable across different regions/nations and value types (see Figure~\ref{results:individualbars}), which likely reflects the conservative nature of the simplex weights in comparison to the optimized regression coefficients. In terms of sharpness, the stacked forecasts were always outperformed by the tight predictions of some of the individual models, and by the EMOS method. 


It is worth noting the delay in the response of the combination methods to changes in individual model performance. For example, the predictions from model eight for hospital bed occupancy in the Midlands improved considerably to become the top performing model on 7th July. The weight assigned to this model, for example in the stacking with time-invariant weights algorithm, increased from $15\%$ on 7th July to $48\%$ on the 14th July, but only became the highest weighted model ($87\%$) for the forecast window beginning on 21st July. This behaviour arises from the requirement for sufficient training data from the improved model to become available in order to detect the change in performance.

\section{Discussion}\label{discussion}


Scoring rules are a natural way of constructing weights for forecast combination. In comparison to Bayesian model averaging, weights obtained from scoring rules are directly tailored to approximate a predictive distribution, and reduce sensitivity to the choice of prior distribution. Crucially use of scoring rules avoids the pitfall associated with model averaging of convergence to a predictive density from a single model, even when the ensemble of models does not include the true data generator. Guidance is available in the literature for which situations different averaging and ensemble methods are appropriate \citet{hoge2019}.
 
In this study, several methods (stacking, EMOS and QRA) to combine epidemiological forecasts have been investigated within the context of delivery of scientific advice to decision makers during a pandemic. Their performance was evaluated using the well-established {\it sharpness}, {\it bias} and {\it calibration} metrics as well as the interval score. When averaged over nations/regions, the best performing forecasts according to both the metrics and interval score, originated from the time-invariant weights stacking method for new daily deaths, EMOS for ICU occupancy, QRA for hospital bed occupancy. However, the performance metrics for each model and ensemble method were found to vary considerably over the different regions and value type combinations. Whilst some individual models were observed to perform consistently well for particular region and value type combinations, the extent to which the best performing models remain stable over time requires further investigation using additional forecasting data.

The rapid evolution of the models (through changes in both parameterization and structure) during the COVID-19 outbreak has led to substantial changes in each model's predictive performance over time. This represents a significant challenge for ensemble methods that essentially use a model's past performance to predict its future utility, and has resulted in cases where the ensemble forecasts do not represent an improvement to the individual model forecasts. For the stacking approaches, this challenge could be overcome by either (a) additionally providing quantile predictions from the latest version of the models for (but not fit to) data points within a training window, or (b) sampled trajectories from the posterior predictive distribution for the latest model at data points that have been used for parameter estimation. Option (a) would allow direct application of the current algorithms to the latest models but may be complicated by addition of model structure due to, for example, changes in control measures, whilst (b) would enable application of full Bayesian stacking approaches using, for example, leave-one-out cross validation \cite{yao2018using}. However, it is important to consider the practical constraints on the resolution of information available during the rapid delivery of scientific advice during a pandemic. With no additional information, it is possible to make the following simple modification to the regression approaches to reduce the prediction bias associated with changes to the model structure or parameters.

\begin{figure}[H]
	\centering
		\includegraphics[width=0.8\textwidth]{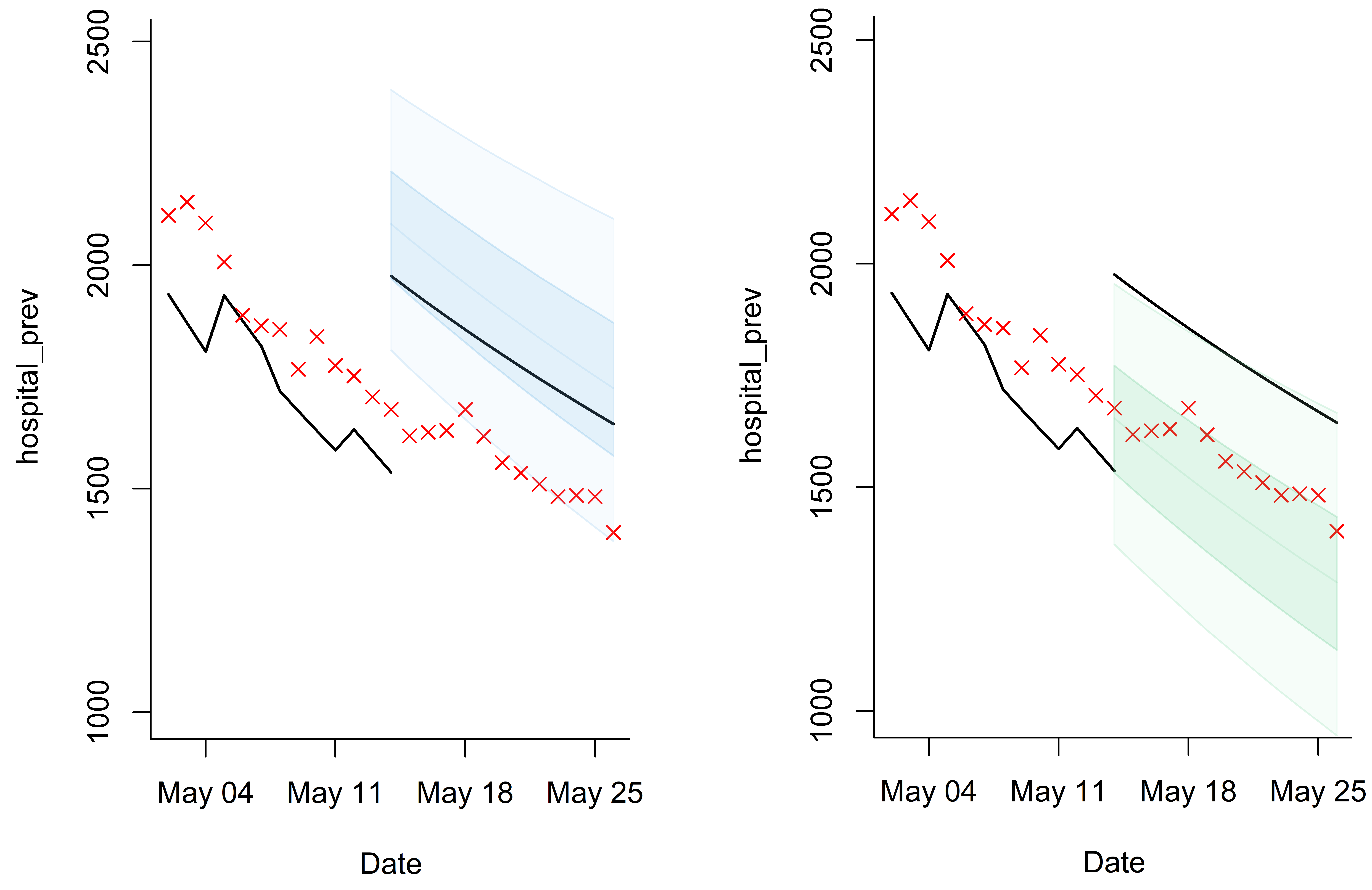}
	\caption{QRA (blue) and Shifted QRA (SQRA; green) forecasts for hospital bed occupancy in the North West region for a forecast window beginning on 14 May. SQRA corrects for the discontinuity between past and current forecasts for the individual model (black line) that corresponds to the covariate with the largest coefficient. Data is shown in red.}
	\label{results:RegressionSolution}
\end{figure}

Analysis of the individual model forecasts revealed large discrepancies between the past and present forecasts for an individual model could lead to increased bias, particularly for the QRA and EMOS combined forecasts. Overlapping of past and present forecasts allows this discrepancy to be characterized, and its impact reduced by translating the individual model predictions (covariates) at time $t$ ($\tilde{y}_{k}(\alpha)$) in the regression models) match the training predictions for the start of the forecast window, $t_0$. For cases where the discrepancy is large (e.g. Figure \ref{results:RegressionProblem}), the reduction in bias of this shifted QRA (SQRA) over QRA is striking, see Figure~\ref{results:RegressionSolution}).

\begin{table}[!htb]
\centering
    \begin{tabular}{llll}
						 & $\bar{S}_{I}$ & $\hat{\beta}$ & $\hat{\gamma}$ \\ \hline
            death\_inc\_line & 1.86 & 0.45 & 0.46 \\
						hospital\_prev & 24.0 & 0.86 & 0.72\\
						icu\_prev  & 2.23 & 0.66 & 0.63\\
        \end{tabular}
		\caption{Median interval, and mean bias and calibration scores for SQRA for each value type, over the four forecast windows.} \label{results:SQRA_table_values}
\end{table}

Table~\ref{results:SQRA_table_values} shows the average interval scores for the SQRA algorithm over the four forecast windows. For new daily deaths and ICU occupancy, the SQRA algorithm achieves better median scores than the other methods considered. These promising results motivate future research into ensemble methods that are robust to the practical limitations imposed by the pandemic response context.

\section*{Acknowledgments}
The research in this paper was undertaken as part of the CrystalCast project, which is developing and implementing methodology to enhance, exploit and visualize scientific models in decision-making processes.

This work has greatly benefited from parallel research led by Sebastian Funk (LSHTM), and was conducted within a wider effort to improve the policy response to COVID-19. The authors would like to thank the SPI-M modelling groups for the data used in this paper.  We would also like to thank Tom Finnie at PHE and the SPI-M secretariat for their helpful discussions.

\vspace{2cm}

(c) Crown copyright (2020), Dstl. This material is licensed under the terms of the Open Government License except where otherwise stated. To view this license, visit \href{http://www.nationalarchives.gov.uk/doc/open-government-licence/version/3}{\color{blue}http://www.nationalarchives.gov.uk/doc/open-government-licence/version/3} or write to the Information Policy Team, The National Archives, Kew, London TW9 4DU, or email: \href{mailto:psi@nationalarchives.gsi.gov.uk}{\color{blue}psi@nationalarchives.gsi.gov.uk}


\end{document}